\documentclass[a4paper,10pt]{article}
\usepackage{pslatex}
\usepackage{mathptmx}
\usepackage[dvips]{graphicx}

\begin{document}
\textsf{\textbf{
\begin{center}
  {Williamsburg, Virginia}
  \\
  \vspace*{2pt} {\LARGE{ACTIVE 04}}
  \\
  {2004 September 20-22}
\end{center}
\begin{flushleft}
  {\LARGE{Indoor Sonic Boom Reproduction Using ANC}}
\end{flushleft}
}
\begin{flushleft}
  {Nicolas Epain, Emmanuel Friot, Guy Rabau}
  \\
  {CNRS - Laboratoire de M\'ecanique et d'Acoustique}
  \\
  {31, chemin Joseph Aiguier}
  \\
  {13402 Marseille cedex 20 France}
  \\
  epain@lma.cnrs-mrs.fr
\end{flushleft}
\begin{center}
  \textbf{ABSTRACT}
\end{center}
}
\vspace*{-8pt}
\noindent
{The European programs for development of supersonic air-flights
involve new studies on the human perception of sonic boom. Because
this noise includes high-level components at very low-frequency, the
usual psycho-acoustic tests with headphones are not relevant; instead,
the original sound-field can be reproduced with many loudspeakers in a
small room, but the loudspeakers must be controlled for an accurate
reproduction, both in time and space, in an area large enough to
enclose a listener's head. In this paper, Active Noise Control is
applied to sonic boom reproduction through Boundary Surface Control
(as named by S.Ise) of the acoustic pressure around a listener. A
small room was built at LMA with sixteen powerful low-frequency
acoustic sources in the walls. Frequency and time-domain numerical
simulations of sonic boom reproduction in this room are given,
including a sensitivity study of the coupling between a listener's
head and the incident sonic boom wave which combine into the effective
sound-field to be reproduced.}

\vspace*{10pt}
\noindent
{\textsf{\textbf{1. INTRODUCTION}}}

\noindent
{Most of the time, when the disturbance induced by some industrial or
  transport noises has to be evaluated, psycho-acoustic tests are
  conducted through sound reproduction using headphones. However, in
  some cases, tests with headphones are not relevant because of
  spectral or spatial specificities of the soundfield to be
  reproduced. Sonic boom is such a very special noise: it is very loud
  (120~dB is a common level for this type of sound), and most of its
  energy is localized at very low frequencies, down to 2 or 3~Hz.
  Headphones could reproduce such a pressure at the eardrums of a
  listener, but it is expected that the perception of such a very low
  frequency sound does not only depend on the pressure fluctuation at
  the listener's ear, but also at the whole listener's body,
  especially on his or her torso. Moreover, headphones do not
  replicate accurately the noise at the eardrums if the listener's own
  Head Related Transfer Functions (HRTF) are not included in the audio
  reproduction device. The sound image reproduced using headphones
  also moves with the listener's head, whereas slight rotations of the
  head in front of a fixed source are known to be an important factor
  in the source localization. Therefore, a soundfield reproduction
  technique is needed that works in an area large enough to enclose a
  person and to let him or her move slightly. The reproduction
  technique should also not depend on the physiognomy of the
  listener.}
 
{Because of the above drawbacks of sound reproduction using
  headphones, a Boundary Surface Control (BSC) technique [1], as named
  by S.Ise, has been chosen to perform indoor sonic boom reproduction
  at the LMA. This paper presents the preliminary work that has been
  conducted to that purpose. Firstly, the soundfield reproduction is
  formulated as an Active Noise Control (ANC) problem, and the theory
  underlying Boundary Surface Control is briefly introduced. Numerical
  simulations of soundfield reconstructions are then presented and
  discussed. Finally, the results of a study on the influence of the
  listener's presence upon the system performances are shown.}

\vspace*{10pt}
\noindent
{\textsf{\textbf{2. SOUND REPRODUCTION USING ANC}}}

\noindent
{The simplest ANC set-up includes a single acoustic source cancelling
  the noise measured by a single microphone. If $d$ denotes the
  disturbance signal and $y$ the signal produced by the secondary
  source, the usual ANC problem is the minimization of the error
  signal $e=d+y$. The nearer $y$ is to $-d$, the
  better is the control performance.}
        
{Therefore, cancelling a primary noise amounts to reproducing it and
  inverting its phase. The ANC problem can be transposed into a sound
  reproduction problem simply by denoting $d$ the sound to be
  reproduced and $y$ the reproduced sound, the error being $e=d-y$.
  This remains true whether the control filter is adaptive or not, and
  for the single-channel as well as for the multi-channel case where
  $d$, $y$ and $e$ are signals vectors. This means that any active
  noise control device can perform sound reproduction. All the
  techniques, algorithms and hardware that have been developed for ANC
  can be used for sound reproduction.}
        
{In this paper, the sound reproduction is formulated as an ANC problem
  because of two features of ANC systems which are not present in
  usual reproduction techniques such as stereophony and 5.1 sound
  reproduction, nor in more advanced techniques such as Wave Field
  Synthesis [2]: ANC devices very often rely on the monitoring of error
  signals which directly measures the system performances, and they
  make use of adaptive filtering. These two features are of importance
  for an accurate sound reproduction: adaptive filtering can reduce
  the sound reproduction sensitivity to temperature changes or to a
  listener's presence; the inclusion of error signals in the
  reproduction process means that direct information about noise
  propagation is available. No a priori assumptions have to be made
  about the reproduction area, an accurate and adaptive reproduction
  can be achieved even in a room at low-frequency. Furthermore, the
  prediction of the primary noise, which requires a feedforward
  reference signal for control of broadband non-stationary noise, is
  not a problem in sound reproduction. The signal $d$, which has to be
  reproduced, must have been recorded or computed in advance so that
  it is available as a convenient reference signal. Moreover, it is
  possible to make use of this reference signal with a time advance as
  long as required in order to make causal the inversion of the
  secondary path matrix which is required for an accurate sound
  reproduction.}

\vspace*{10pt}
\noindent
{\textsf{\textbf{3. BOUNDARY SURFACE CONTROL}}}

\noindent
{For psycho-acoustic tests, a soundfield reproduction must be
  performed over a 3D area which must be large enough to surround the
  listener's head. This means that, from one hand, many sensors and
  actuators are required. On the other hand loudspeakers and
  microphones cannot be placed too close to the listener, otherwise
  the tests would be uncomfortable for him or her. The sound
  reproduction quality could also decrease in this case, due to the
  stronger influence of his or her presence on the soundfield to be
  reproduced (see part 5).}

{Fortunately, sound reproduction, as well as noise cancellation, can
  be performed inside a 3D volume by controlling noise only at the boundary surface
  of the volume. To this purpose, Furuya et al. proposed in 1990
  a method called Boundary Pressure Control, and, in the 90's, Ise
  suggested a substitute method called Boundary Surface Control [1].
  Several other versions of the technique have then been presented
  [3]. As it is denoted by its name, BSC aims at controlling the
  pressure in a volume by monitoring noise at its boundary surface.
  It is justified by the Kirchhoff-Helmoltz integral expression of the
  acoustic pressure inside a bounded volume:}

\begin{equation}
p(\textbf{r})=\int \!\!\!\! \int_{\Sigma}
\left[ G(\textbf{s},\textbf{r})\frac{\partial 
       p(\textbf{s})}{\partial \textbf{n}}
      -p(\textbf{s})\frac{\partial
       G(\textbf{s},\textbf{r})}{\partial\textbf{n}}
\right]  \, \textbf{ds}
\end{equation}

\noindent
{where $p(\textbf{r})$ is the sound pressure at a point $\textbf{r}$
  of the volume $\Omega$, $\textbf{s}$ is a point on the surface
  $\Sigma$ of $\Omega$, $\textbf{n}$ the unit vector which is normal
  to the surface and $G$ the Green function of Helmholtz equation in
  free field. This equation shows that pressure at an interior point
  only depends on pressure and its normal derivative on the exterior
  surface. The idea of BSC is therefore to control these values in the
  same way as for multi-channel ANC, e.g. to impose some pressure
  values on a group of error microphones placed all over the boundary
  of the control region.}
        
{The Boundary Pressure Control method relies on the fact that the some
  redundant information can be found in the right hand side of the
  equation (1). Indeed, the limit of equation (1) for $\textbf{r}$
  tending to a point ${\textbf{s}}_{0}$ of $\Sigma$ is:}

\begin{equation}
\frac{1}{2}p({\textbf{s}}_{0})=\int \!\!\!\! \int_{\Sigma}
\left[ G(\textbf{s},{\textbf{s}}_{0})\frac{\partial 
       p(\textbf{s})}{\partial \textbf{n}}
      -p(\textbf{s})\frac{\partial
       G(\textbf{s},{\textbf{s}}_{0})}{\partial\textbf{n}}
\right]  \, \textbf{ds}
\end{equation}

\noindent
{In addition to Eq.~(1), Eq.~(2) shows that surface pressure normal
  derivative can be seen as a function of the pressure on the surface.
  This means that ensuring the right pressure value on $\Sigma$ gives
  the desired soundfield in any point inside the control region.}
        
{Several problems have been raised concerning the BPC and BSC methods.
  The first problem is that the integral in the Kirchhoff-Helmholtz
  equation is a continuous function of the space variable on the
  surface, which implies that the sound pressure at point $\textbf{r}$
  depends on an infinite number of pressure values. Monitoring the
  pressure at a finite number of locations for control implicitly
  relies on the discretization of the integral in equation (1). In
  practice it is not possible to control the pressure at more than a
  few dozen different points, which imposes a limitation on the sound
  reproduction frequency range [4]. A second problem occurs for BPC at
  the eigenfrequencies of the internal Dirichlet problem in $\Omega$,
  where the solution of equation (2) is not unique [3]. This means
  that, at the volume eigenfrequencies, both the acoustic pressure and
  its normal derivative are theoretically needed for the interior
  sounfield to be fully controlled. Only BSC, which a priori requires
  twice as many sensors as BPC, is supposed to work at the
  eigenfrequencies of the inner volume.  However numerical simulations
  showed that a slight dissimetry in a mesh of pressure sensors on the
  boundary surface could be sufficient for ensuring control of the
  noise inside the volume through control of only the pressure at the
  boundary (see Ref.[1] and [3]); the discretization of equation~(2)
  using an irregular mesh could lead to a discretized problem with a
  unique solution. Finally, a few pressure gradient sensors or a few
  pressure sensors in the volume can be used for BPC in addition to
  the pressure sensors, which is the equivalent in the context of ANC
  of the CHIEF method implemented for computation of acoustic fields
  using a Boundary Element Method [5].}
 
{Because no experimental work has been reported on the failure of BPC
  at the inner volume eigenfrequencies, BPC has been chosen to perform
  sonic boom reproduction at LMA. One of the aims of the numerical
  simulations below is to determine if the theoretical control
  singularity at the eigenfrequencies is a real limitation to the use
  of control of pressure only.}

\vspace*{10pt}
\noindent
{\textsf{\textbf{4. NUMERICAL SIMULATIONS OF INDOOR SONIC BOOM
        REPRODUCTION}}}

\noindent 
{In order to reproduce the noise generated on the ground by a
  supersonic aircraft, a reproduction room has been built at the LMA,
  Marseilles, including sixteen powerful low-frequency acoustic
  sources in the walls. The sources were designed to reproduce the
  high low-frequency pressure levels that can be measured in real
  conditions. Each source includes two large loudspeakers driven with
  out-of-phase signals so that the first distorsion harmonic is
  minimized. A simple model of the room and the noise sources has been
  elaborated for numerical simulations of the sound reproduction.}

\begin{figure}[!!t]
 \begin{center} 
  {\includegraphics[width=2.75in]{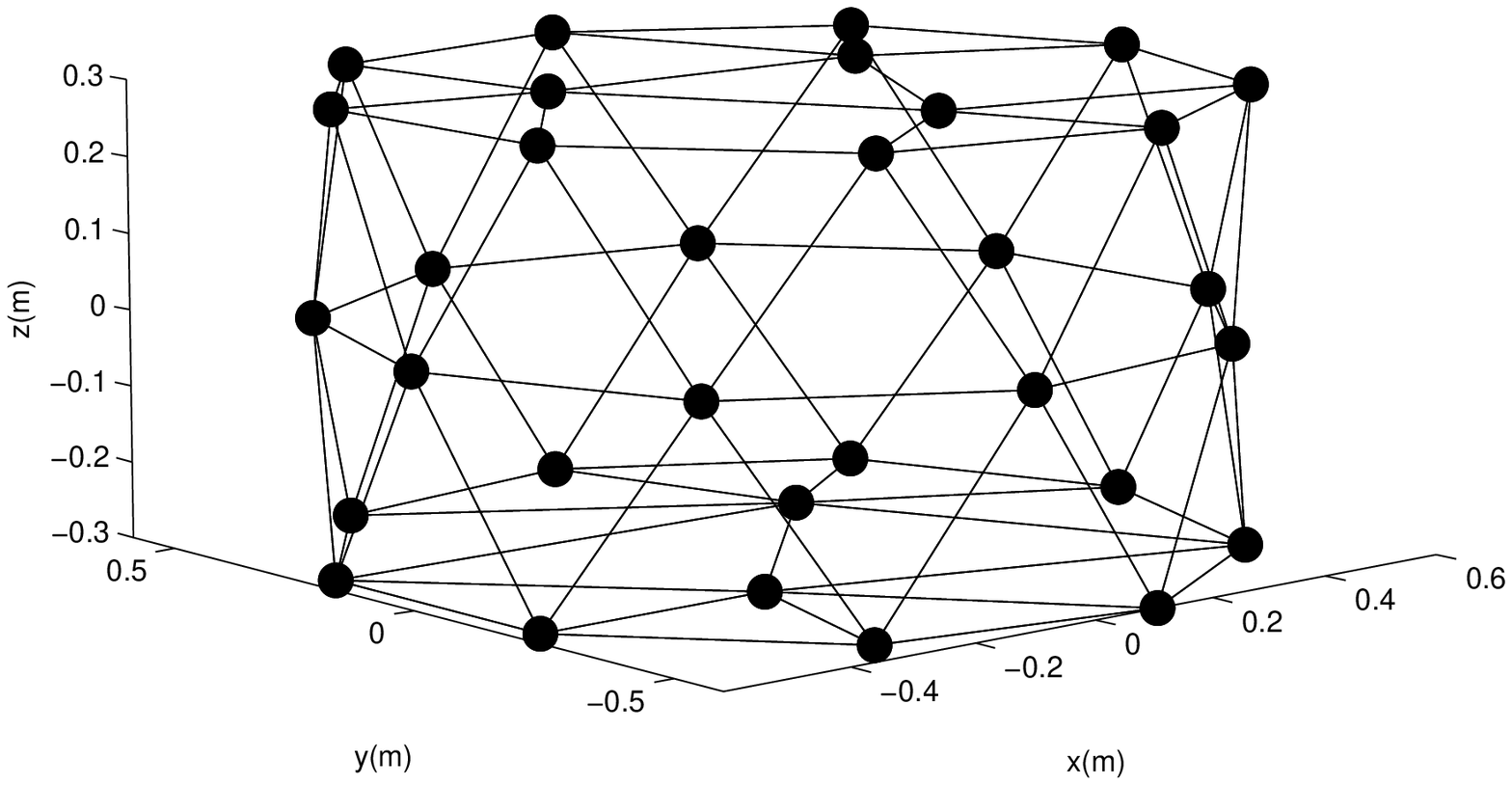} 
   \hspace*{0.5in}
   \includegraphics[width=2.75in]{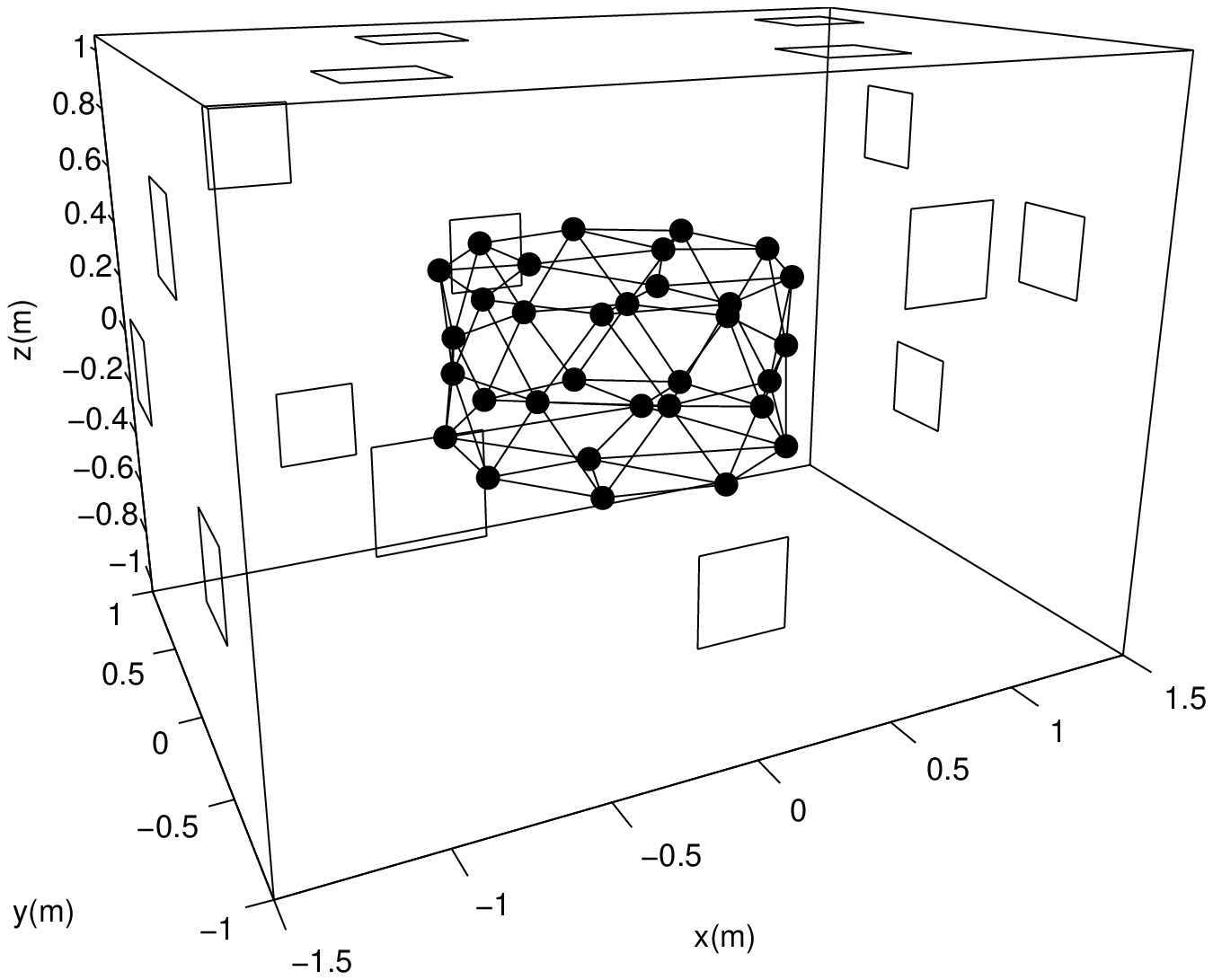}}

  {\textbf{\textsf{
  {a}
  \hspace{3.25in}
  {b}
  }}}
 \end{center}
\caption{The virtual error microphones cylindrical net alone (a), and 
placed into the reproduction room (b). The sources are represented by 
the squares on the walls}
\end{figure}

\vspace*{10pt}
\noindent
{\textsf{\textbf{A. Modelling of the Reproduction Room}}}

\noindent         
{Several simplification hypothesis have been made for modelling the
  reproduction room. The room is assumed to be a perfect 3m x 2m x
  2.1m parallelepiped so that the modal behaviour of the room can be
  easily written as a sum of cosine functions. The absorbing
  properties of the walls are included in the model through a constant
  real normal admittance, which leads to modal damping proportional to
  frequency [4]. Acoustic sources are considered as omnidirectionnal
  monopoles. Finally, it is assumed that the cavity is airtight, i.e.
  there is no acoustic leakage.}
        
{Using these assumptions, the pressure radiated by a monopole source
  at a measurement point can be written as a simple modal series. For
  the simulation this series was restricted to the first thousand
  eigenmodes. This number has been found sufficient to describe the
  acoustical paths in the room at low frequencies (below 500Hz), which
  is the frequency range for which it is intended to reproduce
  accurately the sonic boom soundfield.}
        
{The Boundary Pressure Control method requires the meshing of the
  surface enclosing the volume where sound reproduction is intended.
  For the simulations a cylindrical mesh was considered, which is
  adequate for enclosing a listener during psycho-acoustic tests. The
  cylinder includes 32~microphones/nodes, and has a 60~cm radius and a
  60~cm height, as displayed on figures 1a and 1b. The microphones are
  supposed to be perfectly omnidirectionnal for the simulations.}

\vspace*{10pt}
\noindent
{\textsf{\textbf{B. Frequency-Domain Simulations}}}

\noindent
{In the frequency domain, pressure values and acoustic paths can be
  described by single complex coefficients. The acoustic pressure
  field which has to be reproduced at the error microphones can be
  written as a complex vector ${\textbf{p}}_{M}^{0}$. Let
  ${\textbf{p}}_{P}^{0}$ denote the noise that has to be reproduced
  inside the control volume at some observation points. At a given
  frequency, if ${\textbf{H}}_{SM}$ and ${\textbf{H}}_{SP}$
  respectively denote the transfer matrix from the secondary sources
  to the control microphones, and from the sources to the observation
  point, the optimal vector of source command signal is:}

\begin{equation}
\textbf{q}={\textbf{H}_{SM}^{-1}}{{\textbf{p}}_{M}^{0}} 
\end{equation}

\noindent
{The reproduced soundfield pressure at the observation points is then
  given by:}

\begin{equation}
{\textbf{p}}_{P}={\textbf{H}_{SP}}\textbf{q}={\textbf{H}_{SP}}
{\textbf{H}_{SM}^{-1}}{{\textbf{p}}_{M}^{0}}
\end{equation}

\noindent
{Note that these computations are similar to the derivation of optimal
  noise cancellation: the reconstruction error at the microphones only
  depends on the source-to-microphone transfer matrix inversion.}

\begin{figure}[!!b]
\begin{center}
  \includegraphics[width=2.75in]{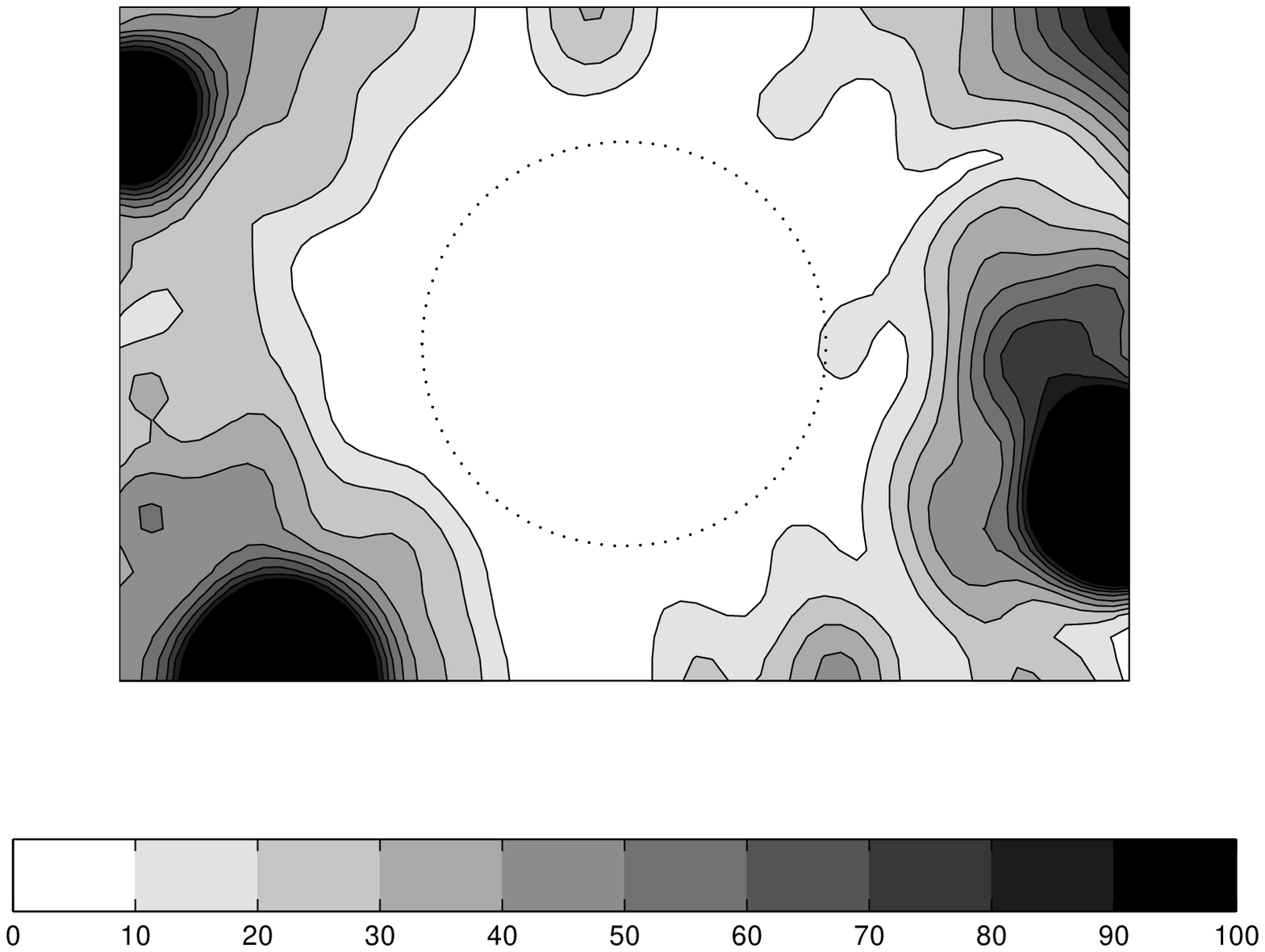}
\end{center}
\vspace*{0.25in}
\begin{center}
  \includegraphics[width=2.75in]{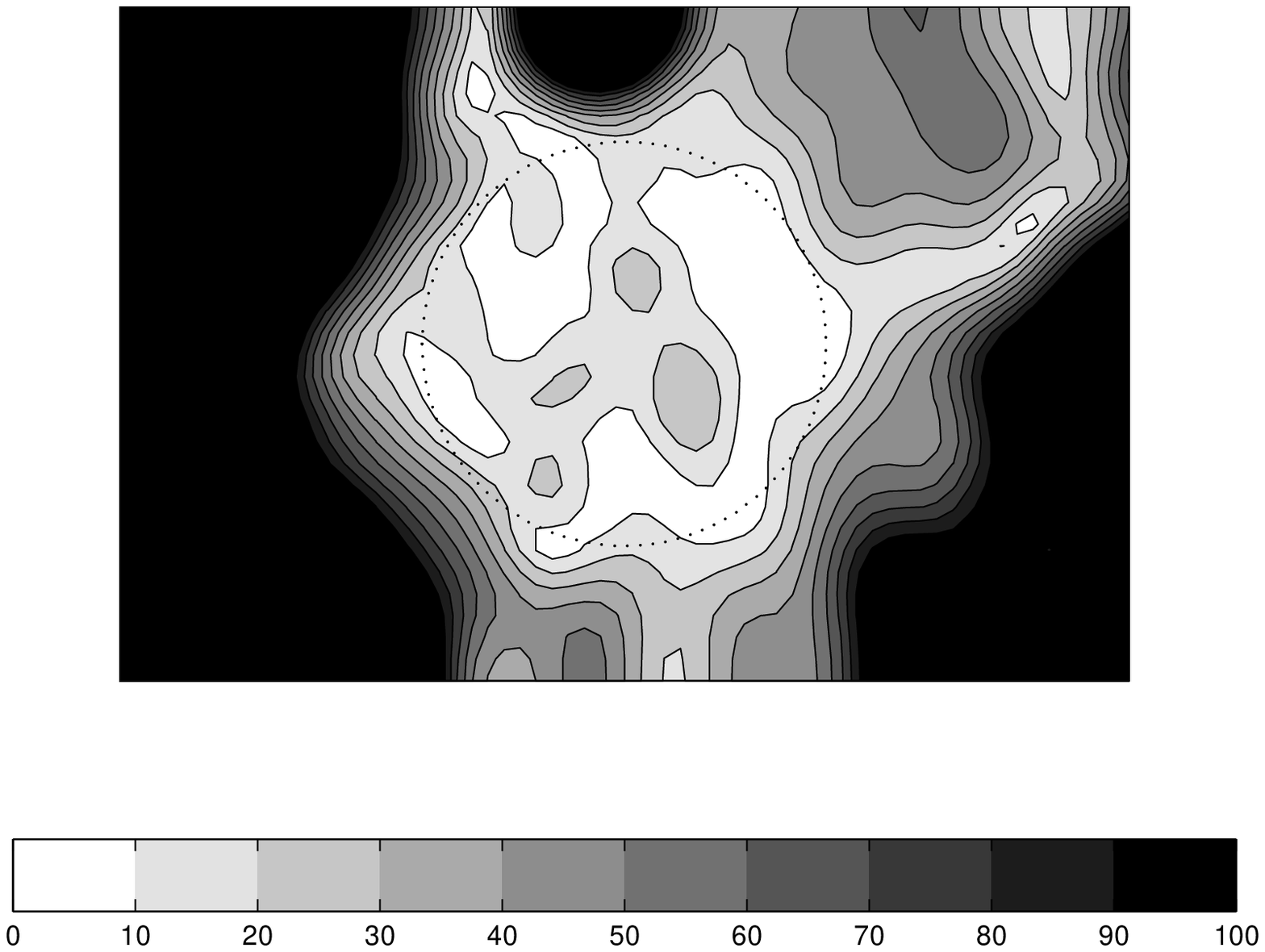}
  \hspace*{0.75in}
  \includegraphics[width=2.75in]{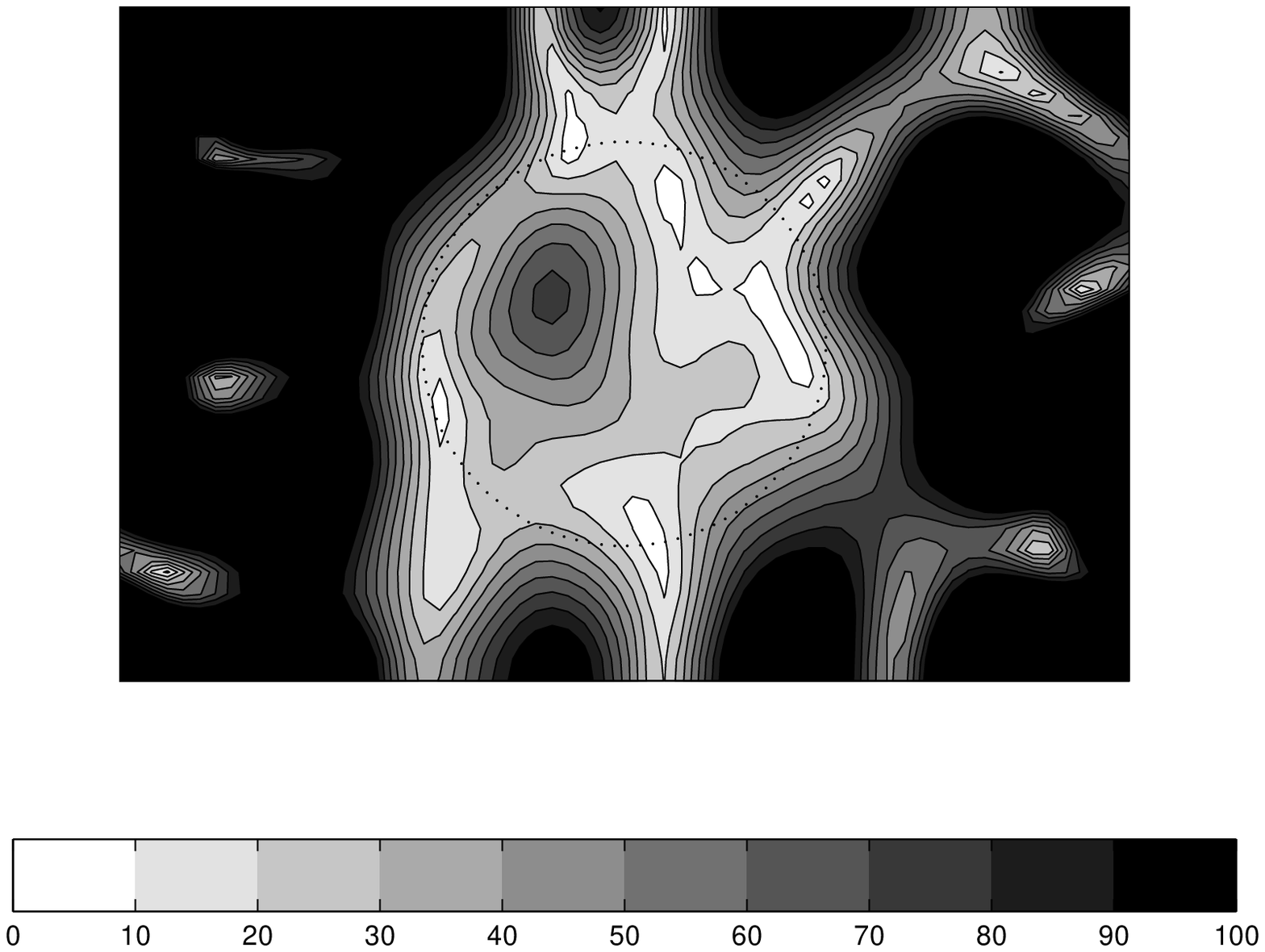}
\end{center}
\caption{Contours of equal sound pressure error level 
  (in percents) for plane waves of 100 Hz (top), 200 Hz (bottom-left)
  and 300 Hz (bottom-right)}
\end{figure}
        
{The first soundfield whose reproduction is evaluated through
  numerical simulations is an harmonic plane wave. In order to provide
  easy-to-interpret figures, an additional horizontal mesh of 32 by 64
  observation points is used to observe the soundfield over the whole
  reproduction room.  The height of the observation points corresponds
  to the middle of the cylinder on the surface of which the acoustic
  pressure is controlled. After computing the input to the control
  source, the pressure error vector can be computed by the following
  formula:}

\begin{equation}
e(x,y)=100\left|
\frac{p_{r}(x,y)-p_{o}(x,y)}{p_{o}(x,y)}\right|
\end{equation}
\noindent
{where $p_{r}$ and $p_{o}$ are respectively the reproduced and
  original pressure values at the $(x,y)$ point. The obtained value
  $e$ is then in percents. Equal sound pressure error contours are
  then displayed for harmonic plane waves at 100, 200 and 300~Hz on
  figure~2. It can be seen that, as expected, the performance of the
  sound reproduction system decreases as the frequency of the original
  sound wave increases. This can easily be explained by the fact that
  the density of microphones per wavelength on the control surface
  decreases as the frequency increases. The
  microphones surface density $D$ is given by:}
 
\begin{equation}
D=\frac{n}{S}
\end{equation}
\noindent
{where $n$ is the number of error microphones and S the surface of the
  cylinder in squared meters. If $S$ is a number of $\lambda$ by
  $\lambda$ squares, where $\lambda$ is the wavelength of the
  original sound, then we have}

\begin{equation}
D_{\lambda}=\frac{n\lambda^{2}}{S}=\frac{nc^{2}}{Sf^{2}}
\end{equation}
\noindent
{where $c$ is the sound speed and $f$ the fequency. The number of
  error microphones for each $\lambda$ by $\lambda$ square decreases
  as a $f^{-2}$ function, and so does the sound reproduction accuracy. This can
  be seen as a generalization of a frequently observed result in ANC,
  which is that three sensors by wavelength are necessary to ensure an
  efficient control along a one-dimensional microphone antenna
  (see Ref.~[4],[6]).}

\begin{figure}[t]
\begin{center}
  \includegraphics[width=100mm]{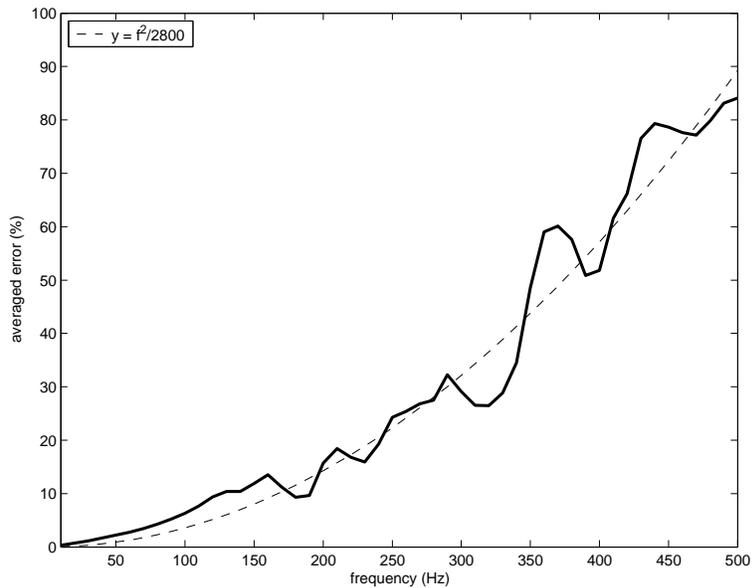}
\end{center}
\caption{Averaged sound pressure error in the control region as a 
function of the frequency}
\end{figure}
         
{In order to observe more accurately the influence of the primary wave
  frequency on the system performance, figure~3 displays the relative
  pressure error, averaged on a few dozens of regularly spaced points
  inside the cylinder where control is intended.
  It can be seen that the reproduction error behaves as
  a $f^{2}$ function, it is inversely proportionnal to the error
  microphones surface density $D_{\lambda}$ of the error microphones,
  which confirms the significance of $D$ as an indicator of the
  reproduction accuracy.}  

{In the case of a cylindrical volume with radius 60~cm and height 60~cm,
  the first eigenfrequency of the Dirichlet problem for the internal
  volume takes place at about 283~Hz, which is within the range of the intended soundfield
  reproduction. It can be seen in figure~3 that no error peak can be
  observed around this frequency, which means that the soundfield
  reproduction through Boundary Pressure Control does not suffer from
  deficiencies at this cylinder eigenfrequency. The eigenfrequencies of higher order
  are out of the frequency range of the system.}

\begin{figure}[!!b]
\begin{center}
  \includegraphics[width=4.7in]{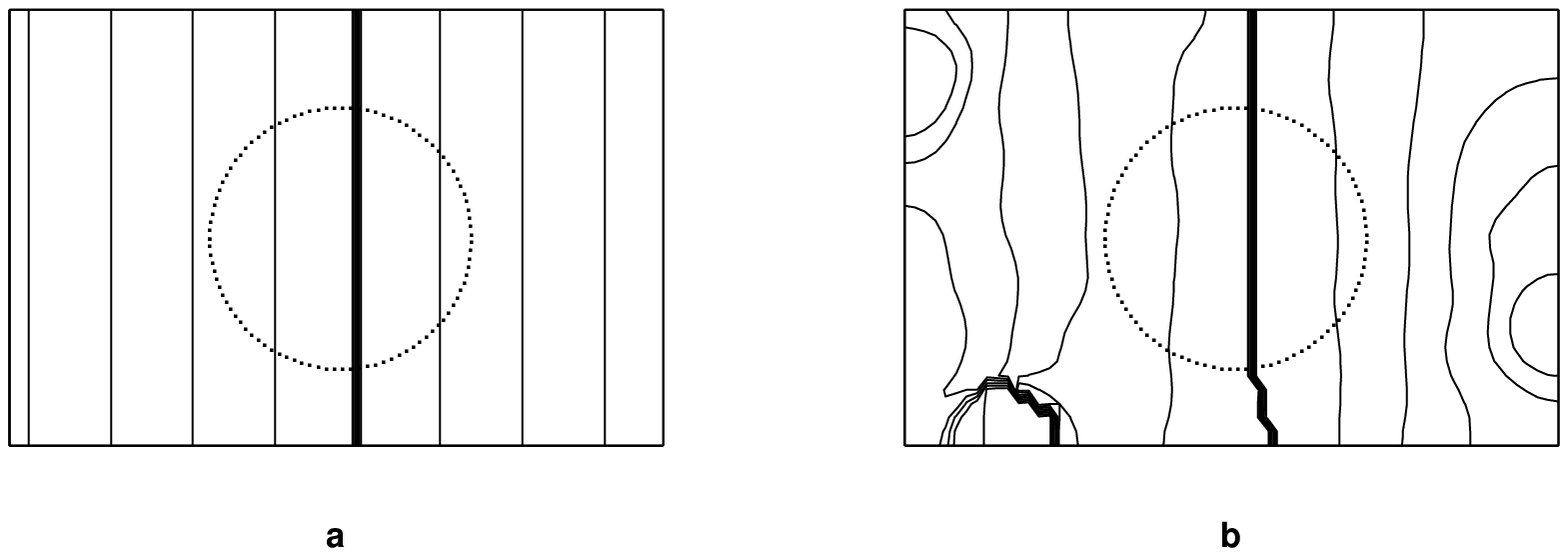}
\end{center}
\vspace*{5pt}
\begin{center}
  \includegraphics[width=4.7in]{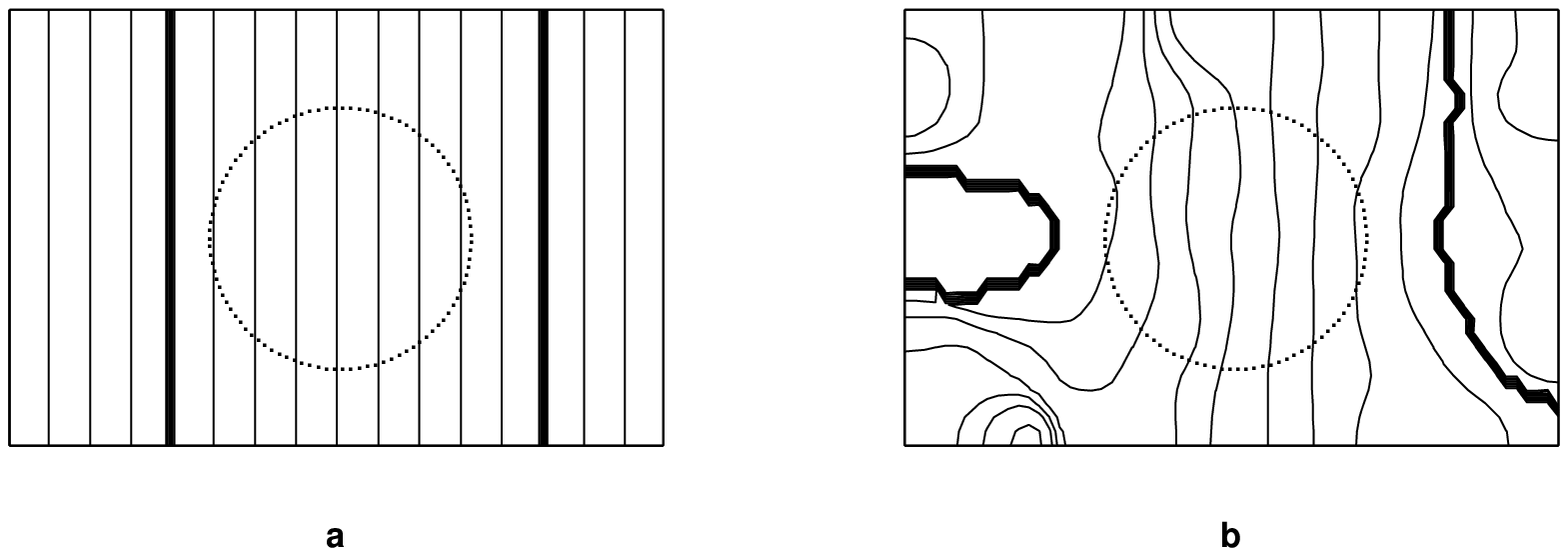}
\end{center}
\vspace*{5pt}
\begin{center}
  \includegraphics[width=4.7in]{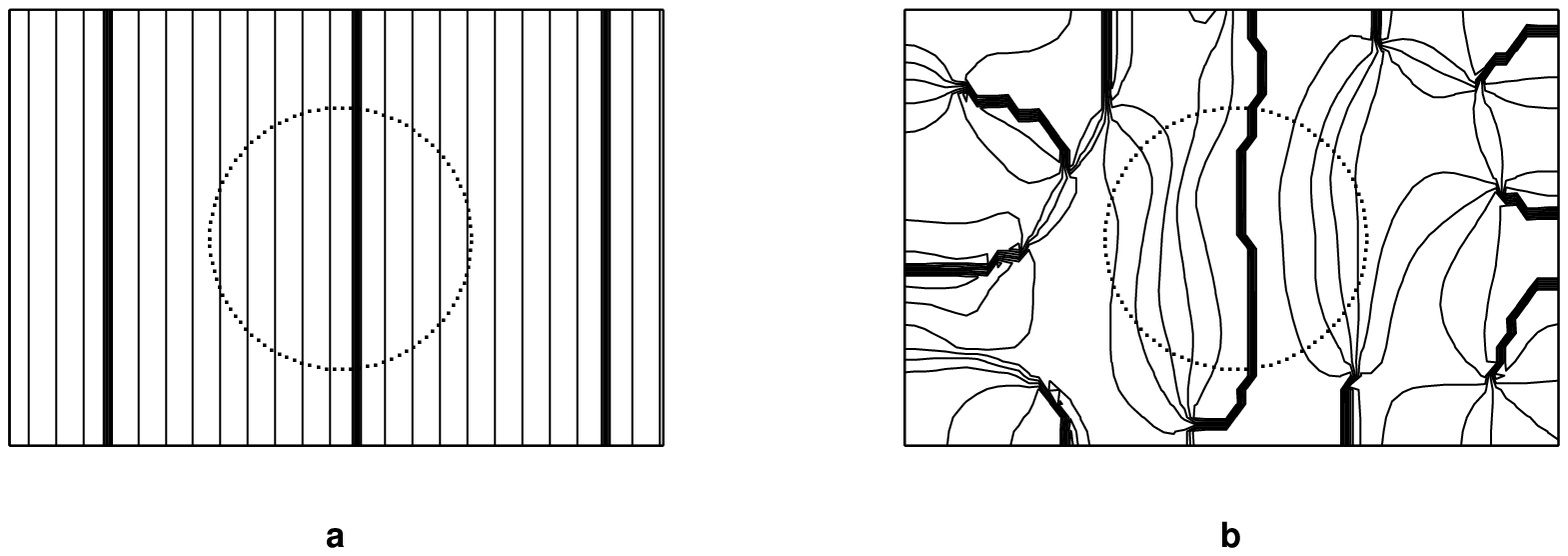}
\end{center}
\caption{Contours of equal sound pressure phase for a 100 Hz 
(top), 200 Hz (middle) and 300 Hz plane wave (bottom); 
a: original soundfield, b: reproduced soundfield)}
\end{figure}

{Although the reproduction error increases quickly with the frequency,
  figures~4 shows that the phase of the secondary wave is quite well
  reproduced even for large values of $f$. This suggests that, even if
  for one frequency value the amplitude of the pressure is not
  perfectly reproduced inside the control region, the crossing of a
  transient sound, for instance from the left to the right, can be
  reproduced so that a listener perceives the direction where the
  sound is coming from, because an accurate phase reconstruction
  involves for the listener a good reproduction of the Interaural Time
  Difference, which is known to be the main cue for the localization
  of noise sources in an horizontal plane at low frequency.}

\begin{figure}[!!!ht]
\begin{center}
  \includegraphics[width=5.45in]{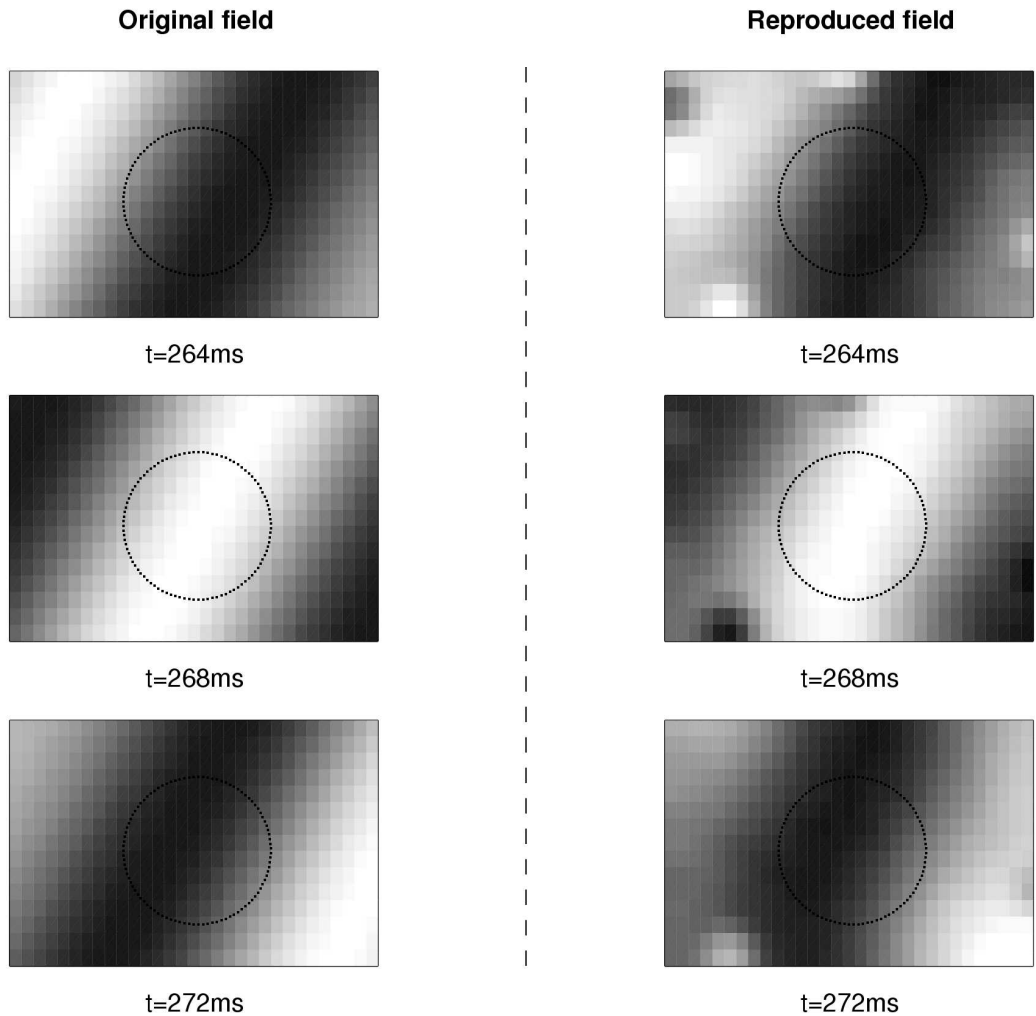}
\end{center}
\caption{Original (left) and reproduced (right) sound pressure for 
several time values. The orginal wave is a plane gaussian pulse with 
center frequency 100 Hz.}
\end{figure}

\begin{figure}[!!!hb]
\begin{center}
  \includegraphics[width=5.45in]{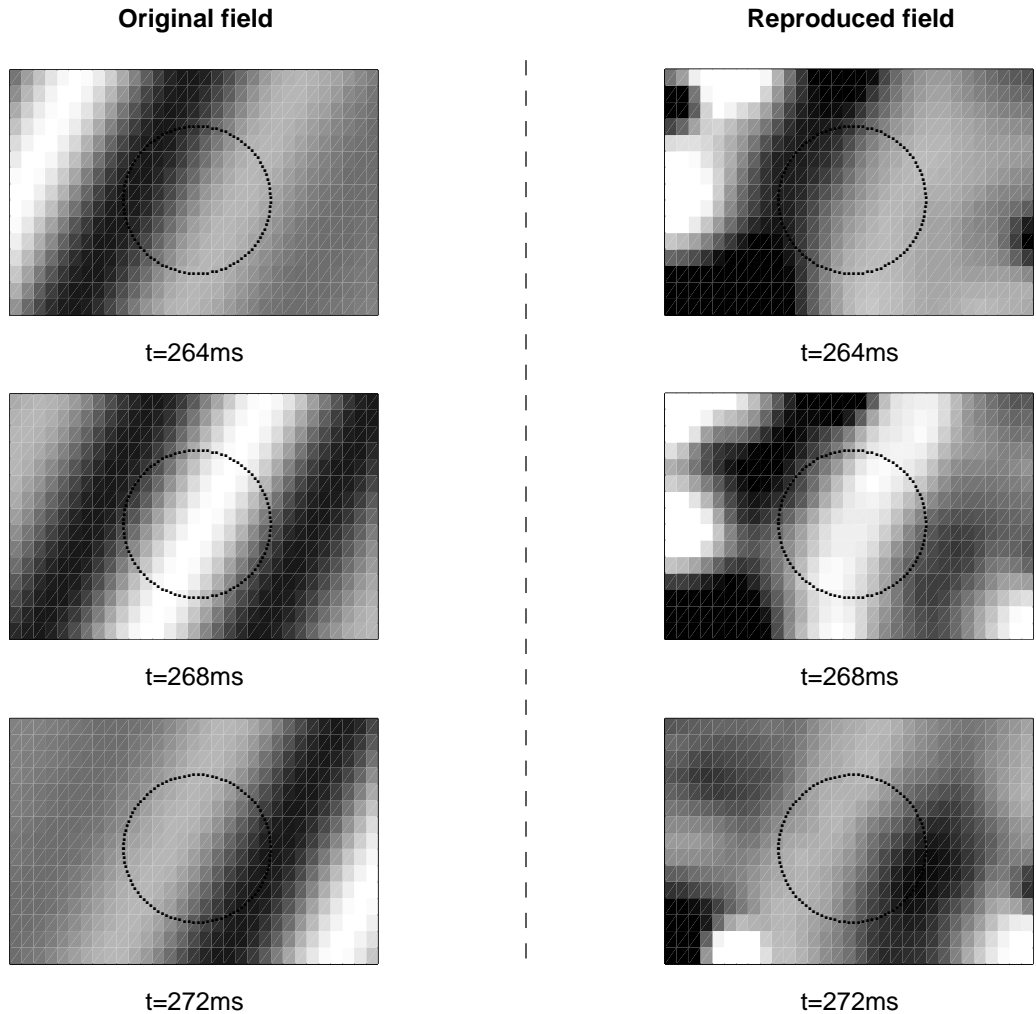}
\end{center}
\caption{Original (left) and reproduced (right) sound pressure for 
several time values. The orginal wave is a plane gaussian pulse with 
center frequency 200 Hz.}
\end{figure}     

\vspace*{10pt}
\noindent
{\textsf{\textbf{C. Time-Domain Simulations}}}

\noindent
{The sonic boom is a very unstationnary noise and, for psycho-acoustic
  tests, the reproduction of the transient noise components must also
  be accurate. Therefore time-domain simulations are required in
  addition to frequency-domain simulations, all the more since
  effective audio reproduction systems work in the time-domain.}
        
{Firstly, the matrix of secondary paths from noise sources to error
  microphones, which has been computed for a dense grid of discrete
  frequencies, gives the corresponding impulse response matrix by using
  Inverse Fast Fourier Transform. The impulse response for the inverse
  of the matrix of secondary paths is also computed through IFFT. Once
  these direct and inverse responses have been obtained, the
  computation of the residual error is made as in the
  frequency-domain, with the difference that the products become
  convolution products. The secondary field is the sum of the
  soundfield from each source, each one being calculated by filtering
  the original signals with the appropriate filters.}

{In a first step, simulations were performed for gaussian pulse plane
  waves, which are the signals with maximum localization in both time
  and frequency. The visualization plane is the same as in the
  previous simulations. This gives, for each time sample, a map of the
  original and reproduced sound pressures in the observation plane,
  which is this time divided in 32 by 16 points. Excerpts of the
  results are presented on figures~5 and 6 for plane gaussian pulse
  waves with center frequency 100 and 200~Hz. The time-domain
  reconstruction of the original wave is very accurate for a 100~Hz
  pulse, and still quite accurate for the 200~Hz one in almost the
  whole controlled region, although the frequency-domain simulations
  suggested a 20 to 30~\% error at this frequency. The results
  observed in the frequency domain for the phase reproduction are
  confirmed, the secondary wavelet travels through the room very like
  the original one.}

\begin{figure}[!!!ht]
\begin{center}
  \includegraphics[width=5in]{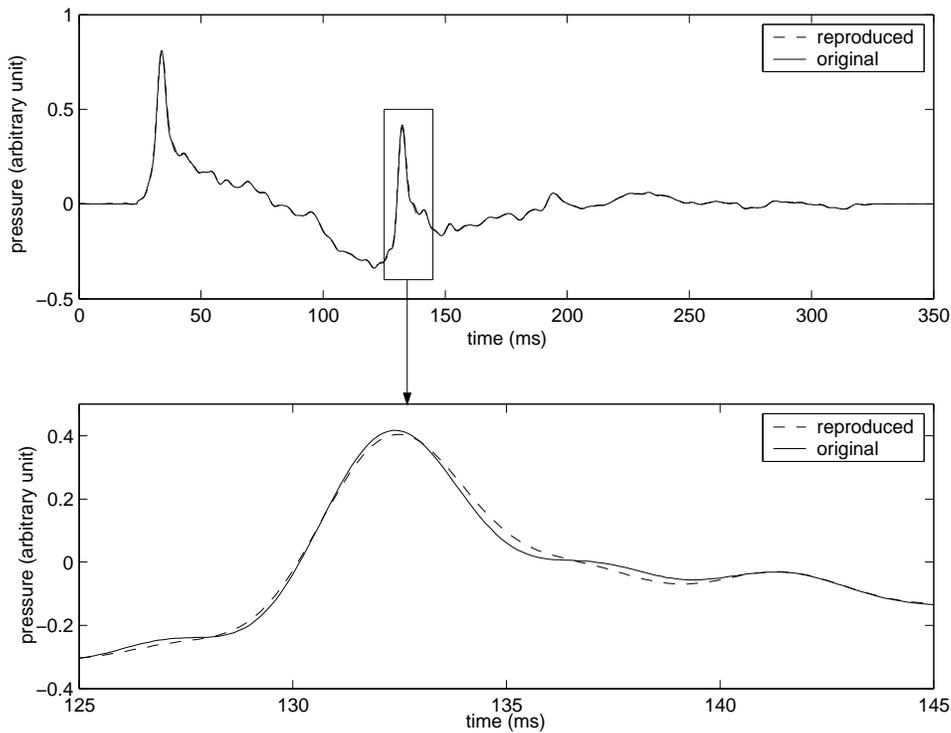}
\end{center}
\caption{Original and reproduced sound pressure for a plane low-pass 
filtered sonic boom wave at the center of the control region.}
\end{figure}  

{In a second step we simulated the reconstruction of a recorded real
  sonic boom. Because of the frequency limitations of the system, the
  recorded sound was low-pass filtered in order to remove the signal
  components at frequencies higher than~300 Hz. The results are
  presented on figure~7 for an observation point placed at the center
  of the control region. Once again, the original soundwave is very
  well reproduced, even if a slight variation of the secondary
  pressure around the original value can be observed. When listening
  the two signals with headphones the difference is almost inaudible.}

\pagebreak[1]
      
\vspace*{10pt}
\noindent
{\textsf{\textbf{D. Conclusions}}}

\noindent
{Frequency and time-domain simulations of indoor soundfield
  reconstruction in the LMA reproduction room have been performed.
  These simulations correspond to an ideal case, where the inverse
  filters are the optimal ones. Thus, the obtained results cannot be
  considered as the real performances of the system in practice but as
  the maximal performances that can be expected. In practice the
  control filters will be Finite Impulse Response filters and the
  reproduced soundfield may not be as close to the original sonic boom
  field as the simulations suggest.}

{However, the simulation results are encouraging enough in
  implementing sonic boom reproduction through Boundary Pressure
  Control. In particular no singular behaviour of the reproduction
  process has been met at the resonance frequencies of the Dirichlet
  eigenproblem for the reproduction area.}

\pagebreak[1]

\vspace*{10pt}
\noindent
{\textsf{\textbf{5. INFLUENCE OF LISTENER'S PRESENCE ON THE
        SOUNDFIELD}}}

\noindent
{When a soundwave, such as a sonic boom one, impings a listener, it is
  diffracted by his or her body, depending on what is the wave
  propagation direction and what are the shape and reflective
  properties of the body. In particular, it is  known that the
  structures of the head and ear pinnae influence the properties of
  the sound measured at the ear drum. Therefore, each listener hears a
  different sound, and the soundfield to reproduce on control
  microphones around a listener by a BPC system is not only the single
  incident sound wave, but the sum of the diffracted and direct waves.
  Thus, an effective sound reproduction requires the
  presence of the listener during both the recording and reproduction
  stages, just as in the case of the use of binaural techniques involving
  HRTF.}
        
{However, since the recording microphones are more distant to the
  listener, his or her influence on the recorded soundfield is expected to be
  lower than in the binaural case. There is therefore a compromise to
  find between a lower listener influence on the soundfield and a
  lower maximum reproduction frequency, because the more distant are
  the microphones from the person, the lower are the surface
  density of sensors and the reproduction accuracy. Moreover, since the
  amplitude of the diffracted pressure depends on the frequency of the
  incident wave (more precisely on the ratio between the wavelength
  and the characteristic dimensions of the diffractive object), it is
  probable that the difference between the direct and undirect
  soundwaves is very tight for low frequencies. For example, for lower
  frequencies under 340~Hz, the wavelength is more than 1~m, whereas a
  common diameter for human head is 17~cm.}
        
{Therefore a few questions need to be answered before implementing BPC
  for reproducing sonic boom around a listener. Firstly, are the
  soundfields recorded around two different persons very different?
  How does this difference depend on the control microphones distance
  and on the incident wave frequency? How accurate is the reproduction
  when the pressure field recorded around a listener A is reproduced
  around a listener B? Finally, is it possible to reconstruct the
  sonic boom around a person by reproducing only the single incident
  soundwave at the control microphones?}

\begin{figure}[!!!b]
 \begin{center}
  {\includegraphics[width=2.125in]{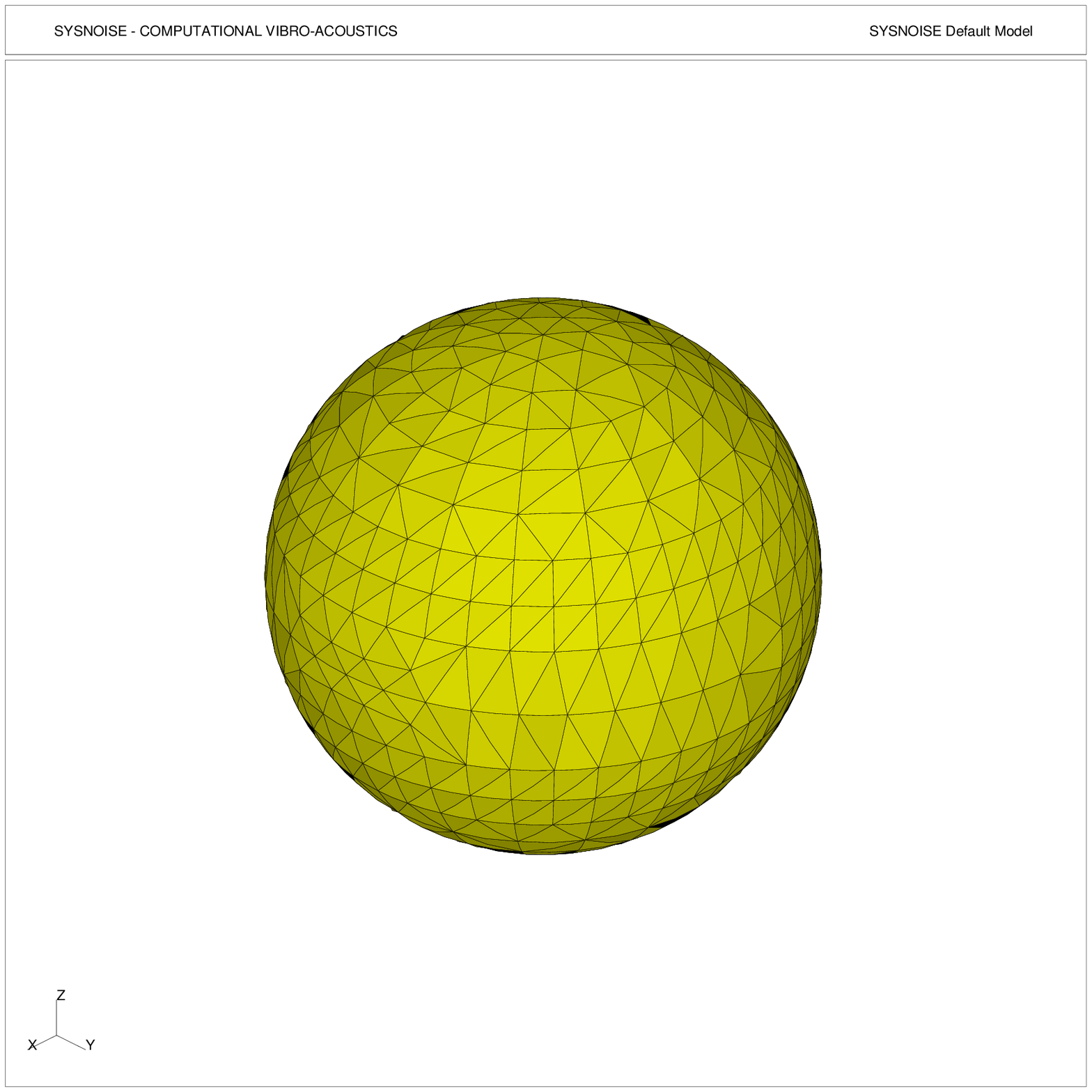}
  \hspace*{1.25in}
  \includegraphics[width=2.125in]{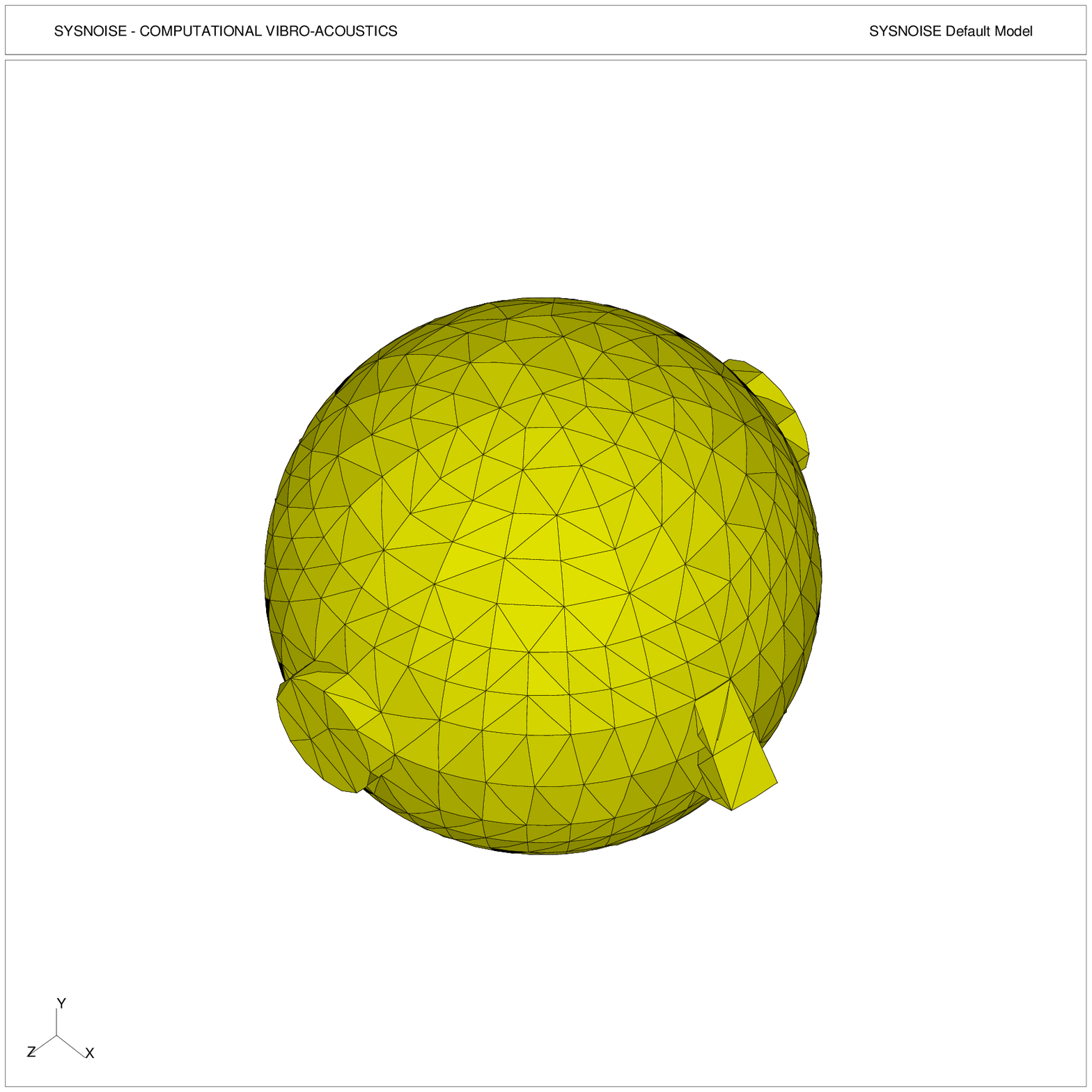}}
 
  {\textbf{\textsf{
   {a}
   \hspace{3.25in}
   {b}
   }}}
 \end{center}
 \caption{The two different finite elements head models used in the 
simulations (a: single rigid sphere; b: sphere with ears and nose).}
\end{figure}  

\begin{figure}[!!!t]
 \begin{center}
  \includegraphics[width=75mm]{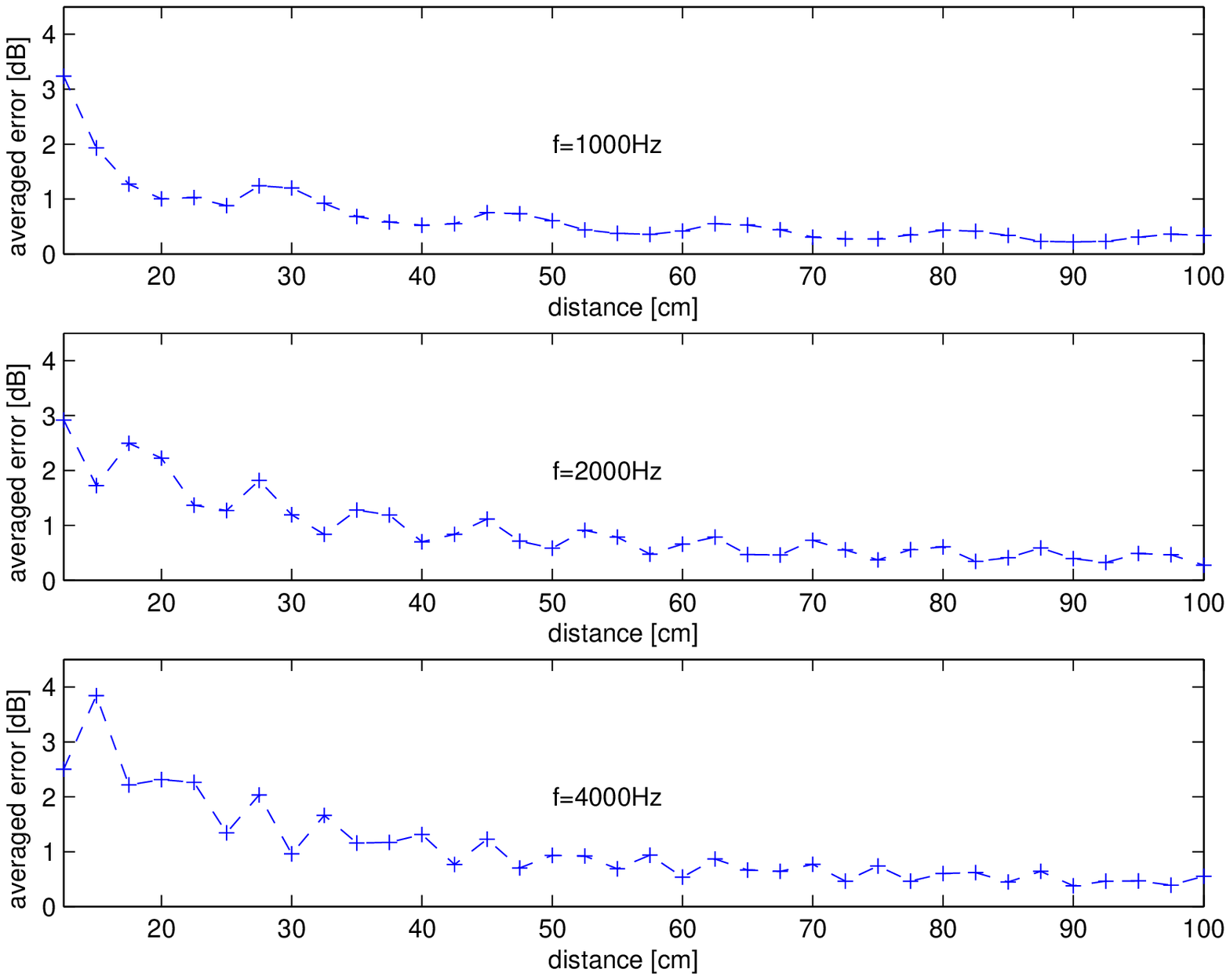}
  \hspace*{9mm}
  \includegraphics[width=78mm]{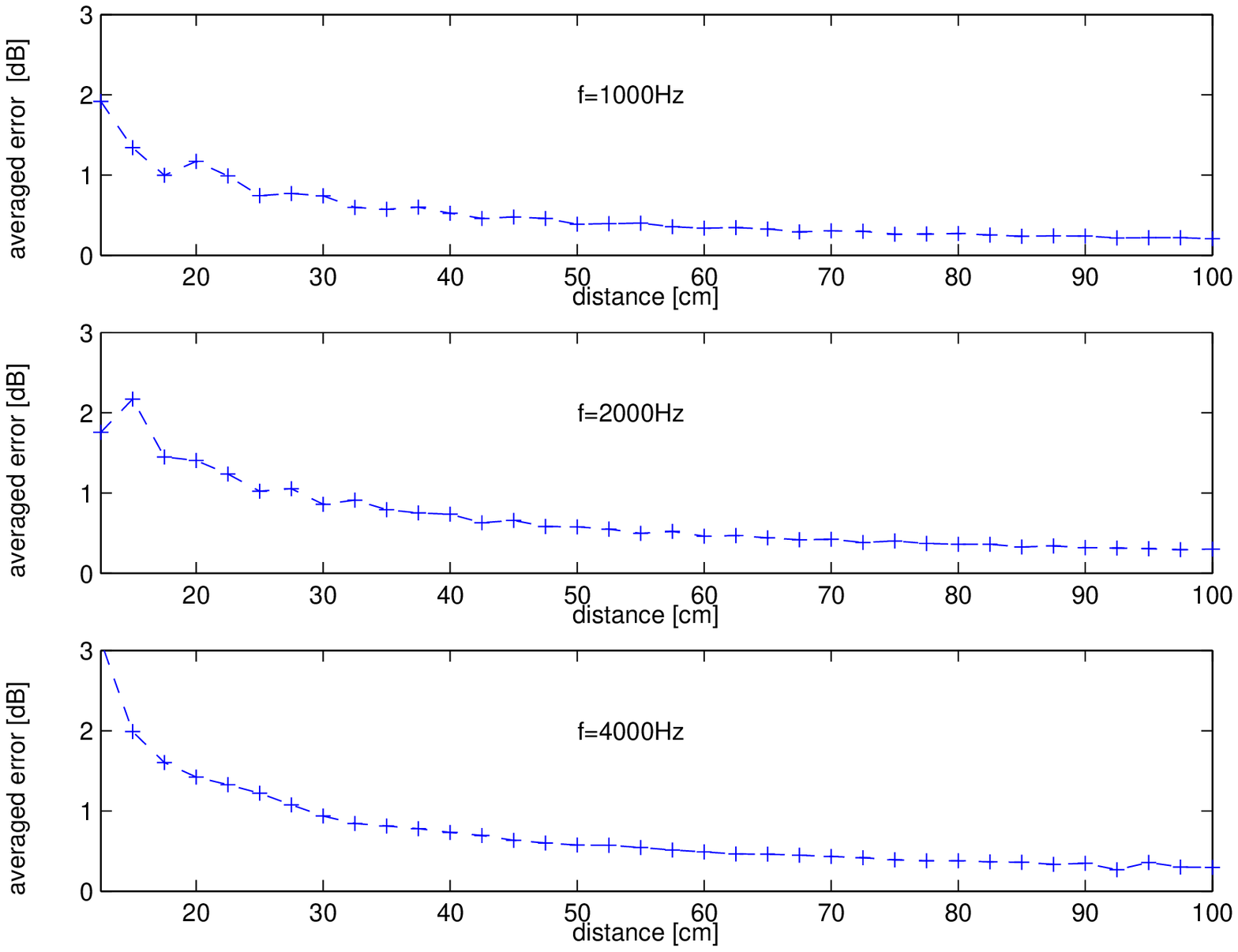}
 \end{center}
 \textbf{\textsf{
  \hspace*{36mm}
  {a}
  \hspace{86mm}
  {b}
 }}
 \caption{Averaged error between: (a) the single incident sound pressure 
and the total pressure around the single sphere; (b) the total pressure 
values around the two head models, as a function of the distance from the 
diffractive object.}
\end{figure}  
  
\vspace*{10pt}
\noindent
{\textsf{\textbf{A. Simulations}}}

\noindent
{In order to answer these different questions, a few numerical
  simulations have been made. Just as before, optimal
  reproduction was computed in the frequency domain for the case of
  plane waves travelling in free-field. Firstly, the
  total field around a rigid sphere (figure 8a) was computed. Figure
  9a shows that the difference between the incident field and the
  total field (including the scattering by the sphere) decreases
  fastly as the distance to the sphere increases. Furthermore
  the decreasing error curve can be divided in two parts: a fast
  decrease occurs bewteen 0 and 50~cm, whereas the decrease is slower
  at more than 50~cm from the sphere. For noise reproduction this
  means that the influence of the scattered noise can be efficiently
  reduced by moving the control microphones away from the sphere
  surface in the 0-50~cm zone, but not that much for larger distances.}

{In a second step the soundfield was computed for a wave impinging a
  finite element head model of a sphere with two ``ears'' and a
  ``nose'' (figure 8b). The difference between this soundfield and the
  soundfield around the mere sphere is shown in figure 9b. As for the
  previous figure, zones of fast and slow decrease appear. Again,
  moving away the microphones from the head of two different listeners
  minimizes the difference in the soundfield that must be reproduced.
  However, this is not so relevant at a distance of more than 50~cm
  when the decrease in the difference is very slow when compared to
  the increase in reproduction error resulting from enlarging the
  controlled area with a constant number of microphones.}

{Finally, optimal reproduction of a low-frequency plane wave
  soundfield in free-field was tested in three configurations: the
  soundfields computed firstly with the sphere, secondly with the head
  model, and thirdly without any scattering object was reproduced by
  BPC around the sphere. It appeared firstly that no large error was
  made when reproducing around the sphere the field computed with the
  head model, at least at low frequencies.  This was expected because
  of the dimensions of the ``ears'' and ``nose'' details compared to
  the wavelength. Secondly, the error resulting from reproducing,
  around the sphere, the noise computed without an object is much
  larger.}

\pagebreak[1]

\vspace*{10pt}
\noindent
{\textsf{\textbf{B. Conclusions}}}

\noindent   
{In conclusion to these different simulations:
\begin{enumerate}
  \setlength{\topsep}{0pt} \setlength{\partopsep}{0pt}
  \setlength{\parskip}{0pt} \setlength{\itemsep}{0pt}
\item {The influence of the diffractive object is reduced at low
    frequencies, i.e. when its dimensions are small when compared to
    the wavelength of the incident sound. It can be noted that this
    will always be true for a human head at frequencies inferior to
    300 Hz.}
\item {At these frequencies, it is possible to use the soundfield
    recorded around a listener A as a reference for the reproduction
    of the soundfield around B.}
\item {It is useless to put the recording/control microphones at a
    distance larger than 50~cm from the listener since the resulting
    benefit in terms of error between the sound pressure values is then
    small when compared to the reproduction accuracy loss which results
    from the enlargement of the control surface.}
\item {Even at low frequency, the reproduction is inaccurate around a
    listener when the field which is reproduced is only made of the
    incident wave (and does not include the scattered wave). This
    shows that it is important to record or compute the original field
    with a person (or perhaps with a dummy head) inside the volume
    defined by the microphones.}
\end{enumerate}}

\vspace*{10pt}
\noindent
{\textsf{\textbf{6. GENERAL CONCLUSIONS AND PERSPECTIVES}}}

\noindent
{Simulations of sonic boom reproduction in a room and, in free field,
  around a human head model, have been performed. Although these
  simulations involve theoretical optimal control, they provide an
  evaluation of the best achievable performances which encourages in
  using BPC for sonic boom reproduction. The LMA room is now ready for
  experiments, sound reproduction and psycho-acoustic tests will be
  conducted soon.}

{It has also been observed numerically that the eigenfrequencies of
  the Dirichlet problem pose no problem for sonic boom reproduction
  with BPC. A free-field experiment of ANC in a volume using the BPC
  method will also be conducted soon at the LMA in order to confirm
  this numerical result.}

\vspace*{10pt}
\noindent
{\textsf{\textbf{7. REFERENCES}}}

\noindent
\begin{enumerate}
  \setlength{\topsep}{0pt} \setlength{\partopsep}{0pt}
  \setlength{\parskip}{0pt} \setlength{\labelsep}{18pt}
  \setlength{\labelwidth}{0pt} \setlength{\leftmargin}{18pt}
  \setlength{\itemsep}{10pt} \vspace*{-6pt}
\item {S. Ise, ``A Principle of Sound Field Control Based on the
    Kirchhoff-Helmholtz Integral Equation and the Theory of Inverse
    Systems,'' \textit{Acta Acustica,} \textbf{85,} 78-87 (1999).}
\item {A. J. Berkhout, D. de Vries, P. Vogel, ``Acoustic Control by
    Wave Field Synthesis'', \textit{J. Acoust. Soc. Am.,} \textbf{93},
    2764-2778 (1993).}
\item {S. Takane, Y. Suzuki, T. Sone, ``A New Method for Global Sound
    Field Reproduction Based on Kirchhoff's Integral Equation,''
    \textit{Acta Acustica,} \textbf{85,} 250-257 (1999).}
\item {P. A. Nelson, S. J. Elliott, \textit{Active Control of Sound}
    (Academic Press, London, 1992)}
\item {H. A. Schenck, ``Improved Integral Formulation for Acoustic
    Radiation Problems'', \textit{J. Acoust. Soc. Am.,} \textbf{44},
    41-58 (1968).}
\item {O. Kirkeby, P. A. Nelson, ``Reproduction of Plane Wave
    Soundfields'', \textit{J. Acoust. Soc. Am.,} \textbf{105}(3),
    1503-1516 (1999).}
\end{enumerate}

\end{document}